\documentstyle[12pt]{article}
\vspace{5mm}

\title{\bf Poisson-Lie sigma models over low dimensional real Poisson-Lie groups }

\vspace{3mm}

\author {  S. Hajizadeh \hspace{-3mm}{ \footnote{ e-mail: s$_{-}$hajizadeh@azaruniv.edu}}\hspace{2mm} {\small and }
A. Rezaei-Aghdam \hspace{-3mm}{ \footnote{Corresponding author.
e-mail:
rezaei-a@azaruniv.edu}} \\
{\small{\em Department of Physics, Faculty of science, Azarbaijan University }}\\
{\small{\em of Tarbiat Moallem , 53714-161, Tabriz, Iran  }}}

\begin{document}
\maketitle

\vspace{5mm}

\begin{abstract}
The Poisson-Lie sigma models over nonsemisimple low dimensional
real Poisson-Lie groups are investigated. We find two sided
models on two, three and some four dimensional Poisson-Lie groups
where the Poisson-Lie sigma models over Poissin-Lie groups $G$ and
its dual $\tilde{G}$ are topological sigma models or BF gauge
models.
\end{abstract}

\newpage

\section{\bf Introduction}
$\; \; \; \;$ Poisson sigma models originally were obtained from
nonlinear extension of classical gauge theory by Ikeda \cite{I};
then independently obtained from generalization of 2d
gravity-Yang-Mills systems by Schaller and Strobl \cite{SS}.
Poisson-sigma model is a 2d topological field theory with Poisson
manifold as target space. These models have played a central role
and given unified structure in the study of two dimensional gauge
theories. They are related to 2d gravity \cite{P}, 2d Yang-Mills
\cite{W1}, topological sigma models \cite{W2}, BF models \cite{BT}
and G/G WZW models \cite{G} as special cases. Poisson sigma model
over Poisson-Lie groups are called Poisson-Lie sigma models; these
models over complex simple Lie groups $G$ in the case that $r$
matrix is factorizable and triangular are studied previously in
\cite{CF},\cite{CFG} respectively.

\smallskip
\goodbreak

Here we study Poisson-Lie sigma models over nonsemisimple low
dimensional real Lie groups. Because for most of the Lie algebras
of these Lie groups, ad-invariant metrics are nondegenerate; hence
here we use other new form of the action for these Poisson-Lie
sigma models. The results are different from general results of
\cite{CF} and \cite{CFG} for some Lie groups. For example, in
\cite{CF} it is shown that the models over dual Lie groups
$\tilde{G}$ for the triangular $r$ matrices are BF gauge models
but these are not true in general for our examples. Furthermore,
we find two sided models (models over $G$ and its dual
$\tilde{G}$) for the case of bi-$r$-matrix bialgebras; such that
the Poisson-Lie sigma models over Lie groups $G$ and its dual
$\tilde{G}$ are topological sigma models or BF gauge models.

\smallskip
\goodbreak

The paper is organized as follows: in section 2 we review Poisson
sigma models and Poisson-Lie sigma models and present new form of
the action for the case of Poisson-Lie sigma models. In section 3
after calculation of $r$-matrices and Poisson structures for two
dimensional Poisson-Lie groups, we obtain two sided topological
sigma models over $G$ and $\tilde{G}$. In section 4 we find
topological sigma models and BF gauge models related to three
dimensional Lie bialgebras\cite{RHR}. The models over $(II.i,V)$
are two sided and the models over Poisson-Lie group $III$ with Lie
bialgebra $(III,III.ii)$ are topological sigma model and also BF
gauge model. Furthermore, we see that there are not two
dimensional Yang-Mills and 2d gravity models over three
dimensional Poisson-Lie groups. In section 5 we present two
examples of four dimensional real Lie bialgebras and find a model
which is two sided and models over $G$ and
$\tilde{G}$ are BF gauge model and also topological sigma model.\\\\

\section{\bf Poisson-Lie sigma models }

\subsection{\bf Review of Poisson sigma models}
 $\; \; \; \; $  Poisson sigma model is a two dimensional topological
  sigma model with the Poisson manifold $M$ as the target space.
 The fields of the model are scalar bosonic fields $x^{\mu}:{\sum}\rightarrow {M}$
 ($x^\mu$ are coordinates of $M$ and $\sum$ is the world sheet) and the fields $A_\mu$
 are 1-forms on $\sum$ with values on $T^{*}M$ so that with coordinates
 $\xi^{\alpha}$ on $\sum$, $A$ can be written as $A = A_{\alpha \mu}d\xi^{\alpha}\wedge dx^{\mu}
 = (A_{\alpha \mu}\partial_{\beta}x^{\mu})d\xi^{\alpha}\wedge d\xi^{\beta}$. In these coordinates
  the action of Poisson-sigma model is given by \cite{SS}
\begin{equation}
S_{top}=\int_{\Sigma} d\xi^{\alpha}\wedge d\xi^{\beta}[A_{\alpha
\mu}(\xi)x^{\mu}_{,\beta}(\xi)+\frac{1}{2}P^{\mu
\nu}(x(\xi))A_{\alpha \mu}(\xi)A_{\beta \nu}(\xi)],
\end{equation}
such that with the notation $A_{\mu}(\xi)=A_{\alpha
\mu}(\xi)d\xi^{\alpha} $, takes the following form:
\begin{equation}
S_{top}=\int_{\Sigma} A_{\mu}\wedge dx^{\mu}+\frac{1}{2}P^{\mu \nu
}A_{\mu}\wedge A_{\nu}.
\end{equation}
The different two dimensional topological models such as
topological sigma models, BF gauge models, two dimensional
gravity, two dimensional Yang-Mills and G/G WZW models are a
special examples of Poisson sigma models. For example if the
Poisson structures on $M$ is nondegenerate (i.e when $M$ is a
symplectic manifolds) then by use of equation of motion for
$A_\mu$ we have
\begin{equation}
A_\mu=\Omega_{\mu \nu}dx^{\mu},
\end{equation}
where $\Omega_{\mu \nu}$ is the symplectic structure (inverse of
$P^{\mu \nu}$). Then in this case the Poisson-sigma model is a
Witten's topological sigma models \cite{W2} with the following
action:
\begin{equation}
S=\int_{\sum} \Omega_{\mu \nu}dx^{\mu}\wedge dx^\nu.
\end{equation}
On the other hand, if Poisson structures are linear in terms of
coordinate $M$, i.e $P^{\mu \nu}=f_{\lambda}^{\mu
\nu}x^{\lambda}$ where $f_{\lambda}^{\mu \nu}$ is the structure
constant of some Lie algebras $\mathbf{g}$ such that its dual
$\mathbf{g}^*$ is identified with $M$\cite{SS};  then the action
(1) can be rewritten as the form of a BF gauge theory
\begin{equation}
S=\int_{\Sigma} x^{\lambda}(dA_{\lambda}+\frac{1}{2}f^{\mu
\nu}_{\lambda}A_{\mu}\wedge A_{\nu})=\int_{\Sigma}
x^{\lambda}F_{\lambda},
\end{equation}
where $F_\lambda$ is the standard curvature 2-form of the gauge
group associated to Lie algebra $\mathbf{g}$. Furthermore, by
choosing $C$ as a quadratic casimir of $\mathbf{g}$ then the
following action:
\begin{equation}
S=S_{BF}+\int C(x(\xi)),
\end{equation}
may be seen to provide the first order formulation of a 2d
Yang-Mills theory \cite{SS}.

To obtain the two dimensional gravity it is enough to choose
Poisson structure as a nonlinear Poisson tensor on $M$ \cite{SS}.
By choosing
$$
P^{\mu \nu}=\varepsilon^{\mu \nu \lambda}u_{\lambda}(x),
$$
$$
u_{a}=\eta_{ab}x^{b},\hspace {1cm}a,b\in\{1,2\},
$$
\begin{equation}
u_{3}=V(x^{a}\eta_{ab}x^{b},x^{3}),\hspace {1cm}\eta_{ab}\equiv
diag(1,\pm 1),
\end{equation}
and inserting these in the action (1) we have
\begin{equation}
S=\int x_{c}(de^{c}+\varepsilon_{ab}\eta^{ac}e^{b}\wedge \omega)+
x^{3}d\omega+V\varepsilon_{ab}e^{a}\wedge e^{b},
\end{equation}
where the first two 1-forms $A_a$ are interpreted as the zweibein
$e_a$ and the third one ($A_3$) as the spin connection $\omega$
of a two dimensional gravity theory. In \cite{ASS} its shown that
gauged G/G WZW models for semisimple groups $G$ are equivalent to
a Poisson-sigma model with a topological term (if Gauss
decomposition of $G$ is complete then the topological term is not
needed).
\\\\

\subsection{\bf Poisson-Lie sigma models }
$\; \; \; \;$ Now, when the Poisson manifold is a Lie group (i.e
$G$ is a Poisson-Lie group) then the Poisson-Sigma model (2) can
be rewritten as the Poisson-Lie sigma model. In \cite{CF} the
action of this model is written as follow:
\begin{equation}
S_1=\int_{\Sigma} (\langle dgg^{-1}\stackrel{\wedge}{,} A \rangle
-\frac{1}{2}\langle A\stackrel{\wedge}{,} (r-Ad_{g}rAd_{g})A \rangle),
\end{equation}
where $g\in G$, $A=A_{\alpha}^{i}d\xi^{\alpha}X_{i}$ and $r\in
\mathbf{g}\otimes \mathbf{g}$ is a classical $r$ matrix with
$\mathbf{g}$ as the Lie algebra of $G$ and $\{X^i\}$ as a basis of
$\mathbf{g}$. Note that the above action can be applied for simple
or nonsemisimple Lie group $G$ with ad-invariant symmetric
bilinear nondegenerate form $\langle X_{i},X_{j}\rangle=G_{ij}$
on the Lie algebra $\mathbf{g}$. When the metric $G_{ij}$ of Lie
algebra is degenerate then the above action is not good. Here we
use the following action instead of the above one\footnote{Note
that the indices in action (2) are geometrical coordinate indices
while in action (10) are Lie algebraic indices.}:
\begin{equation}
S_2=\int_{\Sigma} (dx^{i}\wedge
A_{i}-\frac{1}{2}P^{ij}A_{i}\wedge A_{j}),
\end{equation}
where $x^i$ are Lie group parameters with parametrization (e.g)
\begin{equation}
\qquad \forall g \in G ,\hspace{1cm}
g=e^{x^{1}X_{1}}e^{x^{2}X_{2}}...,
\end{equation}
and $P^{ij}$ is the Poisson structure on Lie group which for
coboundary Poisson-Lie groups it is obtained from Sklyanin bracket
as follows:
\begin{equation}
\{f_1 , f_2\} = \sum_{i,j} r^{ij}((X_i^L f_1)\, (X_j^L f_2) -
(X_i^R f_1)\, (X_j^R f_2))   \qquad \forall f_1 , f_2 \in
C^\infty(G),
\end{equation}
where in the above relation $r^{ij}$ is the classical $r$ matrix
i.e $r:\mathbf{g}\rightarrow \mathbf{g}\otimes \mathbf{g}$, that
is the solution of classical or modified Yang-Baxter equation and
$X_{i}^{L}(X_{i}^{R})$ are left(right) invariant vector fields.
Note that the form of the action (9) in this parameterization can
be rewritten as:
\begin{equation}
S=\int_{\Sigma} R_{m}^{k}[dx^{m}\wedge
A^{i}-\frac{1}{4}(P^{mn}R_{n}^{l}G_{lj}A^{j})\wedge A^{i}]G_{ik},
\end{equation}
where $R_{m}^{k}$ are vielbiens and inverse to the coefficients of
right invariant vector fields. To show the differences between
action (9) and (10) it is better to use an example. For
Poisson-Lie group with Lie bialgebra structure $(V\oplus
R,V.i\oplus R)$ we have\cite{HMRS}(See relations (52-55)):
\begin{equation}
r=X_{1}\wedge X_{4},\hspace {0.5cm}G={\footnotesize
\left(\begin{tabular}{cccc}
                       n & 0 & 0 & m \\
                       0 & 0 & 0 & 0 \\
                       0 & 0 & 0 & 0 \\
                       m & 0 & 0 & p
                     \end{tabular} \right)},\hspace
{0.5cm}P_{12}=x_2 ,\hspace {0.2cm} P_{34}=-x_3,\hspace {0.2cm}
n,m,p\in\Re
\end{equation}
and the form of actions (9) and (10) are obtained as follows:
\begin{equation}
S_1=\int_{\Sigma} ndx^{1}\wedge A^{1}+mdx^{1}\wedge
A^{4}+mdx^{4}\wedge A^{1}+pdx^{4}\wedge A^{4},
\end{equation}
\begin{equation}
S_2=\int_{\Sigma} dx^{1}\wedge A_{1}+dx^{2}\wedge
A_{2}+dx^{3}\wedge A_{3}+dx^{4}\wedge A_{4}-x_{2}A_{1}\wedge
A_{2}+x_{3}A_{3}\wedge A_{4}.
\end{equation}
We see that in this example the Poisson structure does not appear
in the action (9) i.e (15)(because of the effect of degenerate
metric on Lie algebra); and it is not good action for Poisson-Lie
sigma model for this case. In the next sections we use action
(10) to obtain the Poisson-Lie sigma models over all two, three
and some four dimensional Poisson-Lie groups.
\\\\

\section{\bf Models with two dimensional Poisson-Lie groups}
$\; \; \; \;$ The real two dimensional Lie bialgebras are
classified in \cite{HS} as four classes (Abelian, semi-Abelian and
type A and B non-Abelian Lie bialgebras). We find that only type B
non-Abelian Lie bialgebras is coboundary. The commutation
relations for this Lie bialgebra are as follows:
$$
[X_1,X_2]=X_2 ,\hspace {0.5cm}
[\tilde{X}^1,\tilde{X}^2]=\tilde{X}^1 ,
$$
\begin{equation}
[X_1,\tilde{X}^1]=X_2 ,\hspace {0.5cm}
[X_1,\tilde{X}^2]=-X_1-\tilde{X}^2 ,\hspace {0.5cm}
[X_2,\tilde{X}^2]=\tilde{X}^1.
\end{equation}
By use of method \cite{RHR} we find that it is bi-r matrix
bialgebras with $r$ and $\tilde{r}$ matrices and we obtain Poisson
structure as follows:
\begin{equation}
r=X_1\wedge X_2,\hspace
{1cm}\tilde{r}=-\tilde{X}^1\wedge\tilde{X}^2,
\end{equation}
\begin{equation}
P_{12}=\frac{x^1}{1+x^1},\hspace
{1cm}\tilde{P}^{12}=\frac{\tilde{x}_1}{1+\tilde{x}_1}.
\end{equation}
Now the actions $S_2$ and $\tilde{S_2}$ for this models have the
following forms\footnote{Here we use the same parameterization for
$g^*$ as $g$ i.e in relation (11) the tilde parameters and
generators must be replaced with untilde ones.}:
\begin{equation}
S=\int_{\Sigma} [dx^1\wedge A^1+dx^2\wedge
A^2-\frac{x^1}{1+x^1}A^1\wedge A^2],
\end{equation}
\begin{equation}
\tilde{S}=\int_{\Sigma} [d\tilde{x}_1\wedge
\tilde{A}_1+d\tilde{x}_2\wedge
\tilde{A}_2-\frac{\tilde{x}_1}{1+\tilde{x}_1}\tilde{A}_1\wedge
\tilde{A}_2].
\end{equation}
These actions are topological sigma models, because the Poisson
structures are nondegenerate and by use of equation of motions
for $A^i$ i.e
\begin{equation}
\frac{\partial{S}}{\partial{A^1}}=dx^1-\frac{x^1}{1+x^1}A^2=0,
\end{equation}
\begin{equation}
\frac{\partial{S}}{\partial{A^2}}=dx^2-\frac{x^1}{1+x^1}A^1=0,
\end{equation}
and similarly for $\tilde{A}_i$ we can integrate them and obtain
the following actions:
\begin{equation}
S=-\int \frac{1+x^1}{x^1}dx^1\wedge dx^2,
\end{equation}
\begin{equation}
\tilde{S}=-\int_{\Sigma}
{\frac{1+\tilde{x}_1}{\tilde{x}_1}}d\tilde{x}_1\wedge
d\tilde{x}_2.
\end{equation}
Therefore Poisson-Lie sigma models on real two dimensional
Poisson-Lie groups $G$ and its dual $\tilde{G}$ are two sided
topological sigma models. Note that there are two signular points
($x^1=0,x^1=-1$)
for symplectic form $\Omega_{\mu\nu}=\frac{x^1+1}{x^1}$ and its dual.  \\\\

\section{\bf Models with three dimensional Poisson-Lie groups}
$\; \; \; \;$ The real three dimensional Lie bialgebras are
classified in \cite{JR}. Their classical $r$ matrices and their
types (triangular, quasitriangular or factorizable) and also
Poisson structures are obtained in \cite{RHR}. Now by use of
those informations one can obtain Poisson-Lie sigma models over
these Poisson-Lie groups. We perform this work in the following
subsections. \\\\

\vspace{-7mm}

\subsection{\bf Topological Sigma Models}
$\; \; \; \;$ We know that because the number of dimension (three)
is odd then the Poisson structure of three dimensional real
Poisson-Lie groups are degenerate. For this reason one can not
obtain topological sigma models over these Poisson-Lie groups.
But one can obtain topological sigma models over extremal surfaces
(surfaces that specifies from equations of motion) because the
Poisson structures on these surfaces (symplectic leaves) are
nondegenerate. The three dimensional Lie bialgebras that one can
obtain such models over them are $(V,II.i), (III,III.ii),
(III,II), (VI_a,II),$ $(VII_a,II.i)$ and $(VII_a,II.ii)$
\cite{RHR}. In general the Poisson structures for these Lie
bialgebras have the following form:
\begin{equation}
\{x_1,x_2\}=0,\hspace {0.5cm} \{x_2,x_3\}=P_{23},\hspace {0.5cm}
\{x_3,x_1\}=0.
\end{equation}
For these Poisson structures the action $S_2$ has the following
form:
\begin{equation}
S=\int_{\Sigma}[A_{1\alpha}\partial_{\beta}x^{1}+A_{2\alpha}\partial_{\beta}x^{2}+
A_{3\alpha}\partial_{\beta}x^{3}+P^{23}A_{2\alpha}A_{3\beta}]\varepsilon^{\alpha\beta}d{\sigma}d{\tau},
\end{equation}
such that where the equations of motion for $A_i$ are
\begin{equation}
\frac{\partial{S}}{\partial{A_{1\alpha}}}=\partial_{\beta}x^{1}=0,\hspace
{0.5cm}
\frac{\partial{S}}{\partial{A_{2\alpha}}}=\partial_{\beta}x^{2}+P^{23}A_{3\beta}=0,\hspace
{0.5cm}
\frac{\partial{S}}{\partial{A_{3\alpha}}}=\partial_{\beta}x^{2}-P^{23}A_{2\beta}=0.
\end{equation}
From these relations we have
\begin{equation}
x^{1}=cte.,\hspace {0.5cm}
A_{3\beta}=-\frac{1}{P^{23}}\partial_{\beta}x^{2},\hspace {0.5cm}
A_{2\beta}=\frac{1}{P^{23}}\partial_{\beta}x^{3}.
\end{equation}
Then the general form of the action is
$$
S=\int_{\Sigma}[\frac{1}{P^{23}}\partial_{\alpha}x^{3}\partial_{\beta}x^{2}-\frac{1}{P^{23}}
\partial_{\alpha}x^{2}\partial_{\beta}x^{3}-\frac{1}{P^{23}}\partial_{\beta}x^{2}\partial_{\alpha}x^{3}]\varepsilon^{\alpha\beta}d{\sigma}d{\tau}
$$
\begin{equation}
=-\int_{\Sigma}
\frac{1}{P^{23}}\partial_{\alpha}x^{2}\partial_{\beta}x^{3}
\varepsilon^{\alpha\beta}d{\sigma}d{\tau}.
\end{equation}
This model is the topological sigma model over symplectic leaf
$x^1=cte$. Now, for the above mentioned Lie bialgebras we have
the following Poisson structures \cite{RHR} and actions
respectively;

\vspace {0.2cm}

$ (V,II.i):  \{x_1,x_2\}=0 , \hspace {0.5cm}
\{x_2,x_3\}=\frac{1}{2}(e^{2x_1}-1) ,\hspace {0.5cm}
\{x_3,x_1\}=0, $
\begin{equation}
S=-\int_{\Sigma}
\frac{2}{e^{2x_1}-1}\partial_{\alpha}x^{2}\partial_{\beta}x^{3}
\varepsilon^{\alpha\beta}d{\sigma}d{\tau}
   \hspace {1cm}(x_{1}=cte.),
\end{equation}
$ (III,III.ii): \{x_1,x_2\}=0 ,\hspace {0.5cm}
\{x_2,x_3\}=x_{2}+x_{3} ,\hspace {0.5cm} \{x_3,x_1\}=0, $
\begin{equation}
S=-\int_{\Sigma}
\frac{1}{x_{2}+x_{3}}\partial_{\alpha}x^{2}\partial_{\beta}x^{3}
\varepsilon^{\alpha\beta}d{\sigma}d{\tau}
   \hspace {1cm}(x_1=cte.),
\end{equation}
$ (III,II): \{x_1,x_2\}=0 ,\hspace {0.5cm}
\{x_2,x_3\}=\frac{1}{2}(e^{2x_1}-1) ,\hspace {0.5cm}
\{x_3,x_1\}=0, $
\begin{equation}
S=-\int_{\Sigma}
\frac{2}{e^{2x_1}-1}\partial_{\alpha}x^{2}\partial_{\beta}x^{3}
\varepsilon^{\alpha\beta}d{\sigma}d{\tau}
   \hspace {1cm}(x_1=cte.),
\end{equation}
$ (VI_a,II): \{x_1,x_2\}=0 ,\hspace {0.5cm}
\{x_2,x_3\}=\frac{1}{2a}(e^{2ax_1}+1) ,\hspace {0.5cm}
\{x_3,x_1\}=0, $
\begin{equation}
S=-\int_{\Sigma}
\frac{2a}{e^{2ax_1}+1}\partial_{\alpha}x^{2}\partial_{\beta}x^{3}
\varepsilon^{\alpha\beta}d{\sigma}d{\tau}
   \hspace {1cm}(x_1=cte.),
\end{equation}
$ (VII_a,II.i): \{x_1,x_2\}=0 ,\hspace {0.5cm}
\{x_2,x_3\}=\frac{1}{2a}(e^{2ax_1}-1) ,\hspace {0.5cm}
\{x_3,x_1\}=0, $
\begin{equation}
S=-\int_{\Sigma}
\frac{2a}{e^{2ax_1}-1}\partial_{\alpha}x^{2}\partial_{\beta}x^{3}
\varepsilon^{\alpha\beta}d{\sigma}d{\tau}
   \hspace {1cm}(x_1=cte.),
\end{equation}
$ (VII_a,II.ii): \{x_1,x_2\}=0 ,\hspace {0.5cm}
\{x_2,x_3\}=-\frac{1}{2a}(e^{2ax_1}-1) ,\hspace {0.5cm}
\{x_3,x_1\}=0, $
\begin{equation}
S=\int_{\Sigma}
\frac{2a}{e^{2ax_1}-1}\partial_{\alpha}x^{2}\partial_{\beta}x^{3}
\varepsilon^{\alpha\beta}d{\sigma}d{\tau}
   \hspace {1cm}(x_1=cte).
\end{equation}
Note that according to \cite{RHR} all Lie bialgebras except
$(V,II.i)$ and $(III,III.ii)$ are triangular Lie bialgebras but
$(V,II.i)$ and $(III,III.ii)$ are bi-r matrix bialgebras. In this
way the model over $(II.i,V)$ is BF gauge model which we will
consider in the following subsections. Furthermore as we see in
the next subsection the model over $(III,III.ii)$ is also BF
gauge model. \\\\

\vspace{-7mm}

\subsection{\bf BF Gauge Models}
$\; \; \; \;$ As mentioned in section 2, to obtain BF gauge
models, the Poisson structure must be linear in $x^i$. Among
three dimensional real Lie bialgebras only the Poisson structures
of Poisson-Lie groups of $(II.i,V)$ and $(III,III.ii)$ have these
forms \cite{RHR}. For Lie bialgebra $(II.i,V)$ we have the
following Poisson structure for its Poisson-Lie group \cite{RHR}

\vspace{1mm}
$(II.i,V):\hspace {0.2cm} \{x_1,x_2\}=-x_2 ,\hspace
{0.2cm} \{x_2,x_3\}=0 ,\hspace {0.2cm} \{x_3,x_1\}=x_3. $
\vspace{1mm}

Then the action $S_2$ has the following BF form:
\begin{equation}
S=\int_{\Sigma} x^{1}\wedge dA_{1}+x^{2}\wedge dA_{2}+x^{3}\wedge
dA_{3}-x^{2}A_{1}\wedge A_{2}+x^{3}A_{3}\wedge
A_{1}\\=\int_{\Sigma}
x^\mu(dA_{\mu}+\frac{1}{2}\tilde{f}^{\lambda \nu}_{\mu}A_{\mu}
\wedge A_{\nu}),
\end{equation}
where $\tilde{f}^{\lambda \nu}_{\mu}$ are the structure constants
of dual Lie algebra $V$ with the following commutation relations:
\begin{equation}
[\tilde{X}^1,\tilde{X}^2]=-\tilde{X}^2 ,\;\;\;
[\tilde{X}^2,\tilde{X}^3]=0 ,\;\;\;
[\tilde{X}^3,\tilde{X}^1]=\tilde{X}^3.
\end{equation}
Similarly for Lie bialgebra $(III,III.ii)$ we have the following
Poisson structure for its Poisson-Lie group \cite{RHR}

\vspace{1mm}
$(III,III.ii): \hspace {0.2cm}\{x_1,x_2\}=0 , \hspace
{0.2cm} \{x_2,x_3\}=x_{2}+x_{3} ,\hspace {0.2cm} \{x_3,x_1\}=0. $
\vspace{1mm}

Then the action $S_2$ has the following BF form:
\begin{equation}
S=\int_{\Sigma} x^{1}\wedge dA_{1}+x^{2}\wedge dA_{2}+x^{3}\wedge
dA_{3}+(x^{2}+x^{3})A_{2}\wedge A_{3}\\=\int_{\Sigma}
x^\mu(dA_{\mu}+\frac{1}{2}\tilde{f}^{\lambda \nu}_{\mu}A_{\mu}
\wedge A_{\nu}),
\end{equation}
where $\tilde{f}^{\lambda \nu}_{\mu}$ are the structure constants
of dual Lie algebra $III.ii$ with the following commutation
relations:
\begin{equation}
[\tilde{X}^1,\tilde{X}^2]=0 ,\;\;\;
[\tilde{X}^2,\tilde{X}^3]=\tilde{X}^2+\tilde{X}^3 ,\;\;\;
[\tilde{X}^3,\tilde{X}^1]=0.
\end{equation}
Note that because the casimir for Lie algebra $V$ and $III.ii$ are
not quadratic \cite{PSWZ} therefore one can not obtain
2-dimensional Yang-Mills models over these Poisson-Lie groups.

Furthermore, one can not obtain 2D gravity models over Poisson-Lie
group $II.i$ because $u_2(x)=u_3$ and $u_3(x)=-x_2$ (i.e
$u_a=\epsilon_{ab}x^b $, $\epsilon_{23}=1$) and by identifying
$A_a=e_a$ and $A_1=\omega$ we find the action
\begin{equation}
S=\int_{\Sigma} [x^c(de_c+e_c\wedge \omega)+x^{1}d\omega],
\end{equation}
but by use of Cartan structure equation $de_c+e_c\wedge \omega=0$
this action is not good.

One can obtain G/G WZW as a Poisson-Lie sigma model over dual
groups $IX$ and $VIII$ for Lie bialgebras
$(V.i|b,VIII),(V.ii|b,VIII),(V.iii,VIII),(V|b,IX)$ with
topological term; by use of the following action \cite{CF}:
\begin{equation}
S=\int_{\Sigma} (\langle dg.g^{-1}\wedge A\rangle
-\frac{1}{2}\langle A \wedge P_{g}^{\sharp}A\rangle),
\end{equation}
where in the above relation ad-invariant metrics only over Lie
algebras $IX$ and $VIII$ are nondegenerate. Note that Lie
bialgebras $(V.i|b,VIII),(V.ii|b,VIII),(V.iii,VIII),(V|b,IX)$ are
non coboundary \cite{RHR} and the Poisson structures
$P_{g}^{\sharp}$ can be obtained by the method mentioned in
\cite{AGMM} and \cite{AH}.\\\\

\section{\bf Models with some four dimensional Poisson-Lie groups}
$\; \; \; \;$ The classification of real four dimensional Lie
bialgebras is under investigation \cite{HMRS}. Here we give two
interesting examples where the Poisson-Lie sigma models over them
are two sided. The real four dimensional Lie algebras are
classified in \cite{PSWZ}. Note that for the following examples
the ad-invariant metrics are nondegenerate, for this reason one
can not use the action (9) and must use of the action in the form
(10).

a) The commutation relations for Lie bialgebras
$(A_{4,1},A_{4,1}.i)$ are as follows\cite{HMRS}:
\begin{equation}
A_{4,1}: \hspace {0.2cm} [X_2,X_4]=X_1,\hspace {0.2cm}
[X_3,X_4]=X_2,
\end{equation}
\begin{equation}
A_{4,1}.i: \hspace {0.2cm}
[\tilde{X}^1,\tilde{X}^2]=-\tilde{X}^3,\hspace {0.2cm}
[\tilde{X}^1,\tilde{X}^3]=-\tilde{X}^4.
\end{equation}
The classical r-matrices and Poisson structures for these Lie
bialgebras are obtained as:
\begin{equation}
r=bX_1 \wedge X_2+X_1 \wedge X_4+X_2 \wedge X_3, \hspace {0.5cm}
\tilde{r}=\tilde{X}^1 \wedge \tilde{X}^4-\tilde{X}^2 \wedge
\tilde{X}^3+l\tilde{X}^3 \wedge \tilde{X}^4 ,
\end{equation}
\begin{equation}
P_{12}=\frac{1}{2}(x^4)^2+x^3, \hspace {0.5cm}
\tilde{P}^{34}=\frac{1}{2} (\tilde{x}_1)^2-\tilde{x}_2.
\end{equation}
The action and equations of motions for the Poisson-Lie sigma
models over Lie groups $A_{4,1}$ and its dual $A_{4,1}.i$ are the
following form:
\begin{equation}
S=\int_{\Sigma} [dx^1 \wedge A^1+dx^2 \wedge A^2+dx^3 \wedge
A^3+dx^4 \wedge A^4-(\frac{1}{2}(x^4)^2+x^3)A^1 \wedge A^2],
\end{equation}
{\footnotesize
$$
\frac{\partial{S}}{\partial{A^{1}}}=dx^{1}-(\frac{1}{2}(x^4)^2+x^3)A^{2}=0,\hspace
{0.0cm}
\frac{\partial{S}}{\partial{A^{2}}}=dx^{2}-(\frac{1}{2}(x^4)^2+x^3)A^{1}=0,\hspace
{0.0cm} \frac{\partial{S}}{\partial{A^{3}}}=dx^{3}=0,\hspace
{0.0cm} \frac{\partial{S}}{\partial{A^{4}}}=dx^{4}=0,
$$}
\begin{equation}
\tilde{S}=\int_{\Sigma} [d\tilde{x}_1 \wedge
\tilde{A}_1+d\tilde{x}_2 \wedge \tilde{A}_2+d\tilde{x}_3 \wedge
\tilde{A}_3+d\tilde{x}_4 \wedge \tilde{A}_4-(\frac{1}{2}
(\tilde{x}_1)^2-\tilde{x}_2)\tilde{A}_3 \wedge \tilde{A}_4],
\end{equation}
{\footnotesize
$$
\frac{\partial{\tilde{S}}}{\partial{\tilde{A}_{1}}}=d\tilde{x}_{1}=0,\hspace
{0.0cm}
\frac{\partial{\tilde{S}}}{\partial{\tilde{A}_{2}}}=d\tilde{x}_{2}=0,\hspace
{0.0cm}
\frac{\partial{\tilde{S}}}{\partial{\tilde{A}_{3}}}=d\tilde{x}_{3}-(\frac{1}{2}
(\tilde{x}_1)^2-\tilde{x}_2)\tilde{A}_{4}=0,\hspace {0.0cm}
\frac{\partial{\tilde{S}}}{\partial{\tilde{A}_{4}}}=d\tilde{x}_{4}-(\frac{1}{2}
(\tilde{x}_1)^2-\tilde{x}_2)\tilde{A}_{3}=0,
$$}
we see that the models are topological sigma models with the
following actions:
\begin{equation}
S=\int_{\Sigma} \frac{1}{\frac{1}{2}(x^4)^2+x^3}dx^1 \wedge dx^2,
\hspace{1cm} (x^{3}=x^{4}=cte.),
\end{equation}
\begin{equation}
\tilde{S}=\int_{\Sigma} \frac{1}{\frac{1}{2}
(\tilde{x}_1)^2-\tilde{x}_2}d\tilde{x}_3 \wedge d\tilde{x}_4,
\hspace{1cm} (\tilde{x}_{1}=\tilde{x}_{2}=cte.).
\end{equation}

b) The commutation relations for Lie bialgebras $(V \oplus R,V.i
\oplus R)$ are as follows\cite{HMRS}:
\begin{equation}
V \oplus R: \hspace {0.2cm} [X_1,X_2]=-X_2,\hspace {0.2cm}
[X_1,X_3]=-X_3,\hspace {7.7cm}
\end{equation}
\begin{equation}
V.i \oplus R: \hspace {0.2cm}
[\tilde{X}^3,\tilde{X}^4]=\tilde{X}^3,\hspace {0.2cm}
[\tilde{X}^2,\tilde{X}^4]=\tilde{X}^2.\hspace {7.7cm}
\end{equation}
The classical r-matrices and Poisson structures for these Lie
bialgebras are obtained as\footnote{Note that the Poisson
structure on the Lie group $V\oplus R$ is nondegenerate ($det
P=-(x^2x^3)^2$) and we have symplectic structure on this Lie
group.}:
\begin{equation}
r=X_1 \wedge X_4, \hspace {0.5cm} \tilde{r}=-\tilde{X}^1 \wedge
\tilde{X}^4,
\end{equation}
\begin{equation}
P_{12}=x^2,P_{34}=-x^3, \hspace {0.5cm}
\tilde{P}^{12}=-\tilde{x}_2.
\end{equation}
The action and equations of motions for the Poisson-Lie sigma
models over Lie groups $V \oplus R$ and its dual $V.i \oplus R$
are the following form:
\begin{equation}
S=\int_{\Sigma} [dx^1 \wedge A^1+dx^2 \wedge A^2+dx^3 \wedge
A^3+dx^4 \wedge A^4-x^{2}A^1 \wedge A^2+x^{3}A^3 \wedge A^4],
\end{equation}
{\footnotesize
$$
\frac{\partial{S}}{\partial{A^{1}}}=dx^{1}+x^{2}A^{2}=0,\hspace
{0.2cm}
\frac{\partial{S}}{\partial{A^{2}}}=dx^{2}-x^{2}A^{1}=0,\hspace
{0.2cm}
\frac{\partial{S}}{\partial{A^{3}}}=dx^{3}-x^{3}A^{4}=0,\hspace
{0.2cm} \frac{\partial{S}}{\partial{A^{4}}}=dx^{4}+x^{3}A^{3}=0,
$$}
\begin{equation}
\tilde{S}=\int_{\Sigma} [d\tilde{x}_1 \wedge
\tilde{A}_1+d\tilde{x}_2 \wedge \tilde{A}_2+d\tilde{x}_3 \wedge
\tilde{A}_3+d\tilde{x}_4 \wedge
\tilde{A}_4+\tilde{x}_{2}\tilde{A}_1 \wedge \tilde{A}_2],
\end{equation}
{\footnotesize
$$
\frac{\partial{\tilde{S}}}{\partial{\tilde{A}_{1}}}=d\tilde{x}_{1}-\tilde{x}_{2}\tilde{A}_{2}=0,\hspace
{1cm}
\frac{\partial{\tilde{S}}}{\partial{\tilde{A}_{2}}}=d\tilde{x}_{2}+\tilde{x}_{2}\tilde{A}_{1}=0,\hspace
{1cm}
\frac{\partial{\tilde{S}}}{\partial{\tilde{A}_{3}}}=d\tilde{x}_{3}=0,\hspace
{1cm}
\frac{\partial{\tilde{S}}}{\partial{\tilde{A}_{4}}}=d\tilde{x}_{4}=0,
$$}
for this case we also see that the models are topological sigma
models with the following actions:
\begin{equation}
S=\int_{\Sigma}\frac{1}{x^2}\;dx^1 \wedge
dx^2-\frac{1}{x^3}\;dx^3 \wedge dx^4,
\end{equation}
\begin{equation}
\tilde{S}=-\int_{\Sigma} \frac{1}{\tilde{x}_{2}}\;d\tilde{x}_1
\wedge d\tilde{x}_2 ,\hspace
{1cm}(\tilde{x}_3=\tilde{x}_{4}=cte.).
\end{equation}
On the other hand because the Poisson structure for this
bialgebra and its dual are linear, one can obtain BF gauge models
as follows:
\begin{equation}
S=\int_{\Sigma} [dx^1 \wedge A^1+dx^2 \wedge A^2+dx^3 \wedge
A^3+dx^4 \wedge A^4-x^{2}A^1 \wedge A^2+x^{3}A^3 \wedge
A^4]=\int_{\Sigma} x^\mu(dA_{\mu}+\frac{1}{2}\tilde{f}^{\lambda
\nu}_{\mu}A_{\mu} \wedge A_{\nu}),
\end{equation}
where $\tilde{f}^{\lambda \nu}_{\mu}$ is the structure constant
of dual Lie algebra $V.i\oplus R$ with the commutation relation
in (52) and the dual action is
\begin{equation}
\tilde{S}=\int_{\Sigma}[d\tilde{x}_1 \wedge
\tilde{A}_1+d\tilde{x}_2 \wedge \tilde{A}_2+d\tilde{x}_3 \wedge
\tilde{A}_3+d\tilde{x}_4 \wedge
\tilde{A}_4+\tilde{x}_{2}\tilde{A}_1 \wedge
\tilde{A}_2]\\=\int_{\Sigma}
\tilde{x}^\mu(d\tilde{A}_{\mu}+\frac{1}{2}f^{\lambda
\nu}_{\mu}\tilde{A}_{\mu} \wedge \tilde{A}_{\nu}).
\end{equation}
where $f^{\lambda \nu}_{\mu}$ is the structure constant of Lie
algebra $V\oplus R$ with the commutation relation in (51).
\\\\

\section{\bf Conclusion}
We have found Poisson-Lie sigma models on two, three and some four
dimensional Poisson-Lie groups. Most of these Lie groups are non
compact, and ad-invariant symmetric metrics over them are
nondegenerate; for this reason one can not apply the action
presented in \cite{CF} and here we apply other action for
construction of Poisson-Lie sigma models over these Lie groups.
Models over bi-r-matrices bialgebras are two sided. The results
of two sided models are given in the following table (these models
all are new.)
\begin{center}
\hspace{3mm}{\footnotesize Two sided Poisson-Lie sigma models over
low dimensional real Lie bialgebras.}
\begin{tabular}{|c|c|c|}  \hline\hline
$({\bf g}, \tilde{\bf g})$       & Models on $\mathbf{g}$ &
Models on $\tilde{\mathbf{g}}$
\\\hline
$type B:(A_{2},A_{2}.i)$      &topological sigma
model& topological sigma model
\\\hline $(II.i,V)$   & BF gauge
model   &topological sigma model over symplectic leaf
\\\hline $(A_{4,1},A_{4,1})$ & topological sigma
model &topological sigma model \\& over symplectic leaf & over
symplectic leaf
\\\hline
$(V \oplus R,V.i \oplus R)$
 &BF gauge
model& BF gauge model\\
 & topological sigma model& topological
sigma model over symplectic leaf
\\\hline
\end{tabular}
\end{center}

\vspace{2cm}

\bigskip
{\bf Acknowledgments}

\vspace{3mm} We would like to thank  F. Darabi for carefully
reading the manuscript and useful comments.

\end{document}